\documentclass[letterpaper]{article} 
\usepackage{aaai25}  
\usepackage{times}  
\usepackage{helvet}  
\usepackage{courier}  
\usepackage[hyphens]{url}  
\usepackage{graphicx} 
\usepackage{natbib}  
\usepackage{caption} 
\usepackage{amsfonts} 
\usepackage{algorithm}
\usepackage{algorithmic}
\usepackage{booktabs} 
\usepackage{amsmath} 
\usepackage{array}
\usepackage{caption}
\usepackage{graphicx}
\usepackage{subcaption}  
\usepackage{siunitx}
\usepackage[table]{xcolor}
\usepackage{worldflags}   
\usepackage{makecell}

\newcommand{\ctry}[2]{\worldflag[width=3.2mm]{#1}\,#2}

\usepackage{newfloat}
\usepackage{listings}
\DeclareCaptionStyle{ruled}{labelfont=normalfont,labelsep=colon,strut=off} 
\floatstyle{ruled}
\newfloat{listing}{tb}{lst}{}
\floatname{listing}{Listing}
\title{Digital Diasporas: How Origin Characteristics and Host–Native Distance Shape Immigrants’ Online Cultural Retention
\thanks{\textcolor{red}{This paper will appear at ICWSM 2026. Please cite the peer-reviewed version.}}}

\author {
    Aparup Khatua\textsuperscript{\rm 1},
    David Jurgens\textsuperscript{\rm 2},
    Ingmar Weber\textsuperscript{\rm 3}
}
\affiliations{
    \textsuperscript{\rm 1}Luleå University of Technology, Sweden\\
    \textsuperscript{\rm 2}University of Michigan, Ann Arbor, USA\\
    \textsuperscript{\rm 3}Saarland Informatics Campus, Saarland University, Germany\\

    aparup.khatua@ltu.se, jurgens@umich.edu, iweber@cs.uni-saarland.de
}

\usepackage{bibentry}

\begin{document}

\maketitle

\begin{abstract}
Immigrants bring unique cultural backgrounds to their host countries. Subsequent interplay of cultures can lead to either a \textit{melting pot}, where immigrants adopt the dominant culture of the host country, or a \textit{mosaic}, where distinct cultural identities coexist. The existing literature primarily focuses on the acculturation of immigrants, specifically the melting pot hypothesis. In contrast, we attempt to identify the antecedents of the mosaic hypothesis or factors that enhance (or diminish) the propensity for cultural retention among immigrants. Based on Facebook advertising data for immigrants from 8 countries residing in the USA, our findings suggest that greater host–native distance is linked to higher online cultural retention, and while origin country context is statistically significant, its impact is generally smaller. 
\end{abstract}

\section{The Interplay of Cultures}
If a Colombian immigrant lives in Canada, he will be  \textit{confronted with the ever-pressing question ``Who am I?" but also ``Am I Canadian?"}, and eventually, he may become acculturated into the new culture by embracing social activities of the host nation such as \textit{playing hockey with Canadian friends, a sport ignored when he lived in Colombia} \cite[p.~855]{cardenas2017understanding}. This acculturation process helps immigrants develop an identity that enables them to get integrated into the unfamiliar terrain of the host country. However, when immigrants move from one country to another, they also carry along their native cultural backgrounds, characterized by unique languages, religions, and societal values \cite{angelini2015life, cardenas2017understanding}.  The consequence can be either \textit{a melting pot or a mosaic}. In a melting pot scenario, immigrants adopt the dominant culture of the host country, which leads to the homogenization of cultural identities. Conversely, some migrants, regardless of the extent to which they embrace the host culture,  maintain their distinct cultural identities, allowing both cultures to coexist like a mosaic that can (potentially) enrich the host country by fostering cultural diversity.

Extant literature on the interplay of cultures has primarily taken a host-country-centric perspective, explored how migrants adapt to their new communities, and assumed that successful integration primarily depends on the efforts of immigrants \cite{stewart2019rock, montreuil2004acculturation,van2013fading}. The underlying assumption of the \textit{melting pot hypothesis} is that over time, this acculturation process would lead to a homogeneous society where the cultural differences would be less pronounced, and our proverbial Colombian immigrant would start playing hockey and consider himself a Canadian. 
Conversely, according to the \textit{mosaic hypothesis}, immigrants would prefer to maintain their distinct cultural identities. Irrespective of whether they are embracing the host culture, they preserve the traditions and practices of their native country and can contribute to the cultural mosaic of the host country. Ideally, this cultural pluralism will enrich the host country by fostering cross-cultural understanding and appreciation. However, the counter-argument is that a cultural mosaic due to immigrant inflows will be a threat to the cultural identity of the host country \cite{khatua2023we}. 

Cultural retention among immigrants can potentially foster a sense of belonging and resilience, as well as help immigrants preserve their identity and promote psychological well-being in the host country \cite{heinonen2005leisure, lillekroken2024food}. While offline integration may depend on the host-country context, such as the presence of diaspora communities, online platforms provide spaces for maintaining connections to home cultures. Therefore, in today’s digital era, staying connected to one’s cultural roots increasingly occurs online. For instance, while immigrants might face challenges in maintaining their cultural practices in offline settings, online platforms often provide an alternative space for cultural interactions, such as following media from their home country. However, to the best of our knowledge, no prior studies have investigated the determinants of this form of online cultural retention. Thus, our {\textbf{Research Questions} are as follows,  

\textit{\textbf{RQ1}:Does online cultural retention solely depend on the characteristics of their country of origin? \textbf{RQ2}: Is it also affected by the various dimensions of the distance between the host and native countries?}

Our paper aims to address the above RQs by probing online Facebook data, widely used by prior studies to investigate various cultural aspects of immigrants \cite{stewart2019rock, khatua2023host, vieira2020using,dubois2018studying}. We consider the Facebook Advertising API to probe the cultural retention of immigrants from eight countries settled in the USA. Facebook API data is not only an innovative proxy for cultural retention but also captures a different aspect of cultural retention. For instance, an immigrant may be outwardly integrated into her offline life but still only listen to music and read news from her home country \cite{barnett2017predicting,gunsoy2020cultural}. In this way, this study focuses more on online cultural retention than on conventional offline cultural retention.

Following prior studies, we have carefully identified a plausible set of antecedents: both country-specific variables (e.g., economic development, political liberties, civil rights, etc.) as well as different dimensions of distance (e.g., geographic, economic, and Hofstede's 6 Dimensions of Cultural differences) between the host and native countries. To capture cultural retention, we have considered the culinary, musical, and recreational preferences of immigrants \cite{dubois2018studying, stewart2019rock,  vieira2020using, vieira2022interplay, obradovich2022expanding}. Our findings suggest that country-specific factors can have mixed effects on cultural retention, but a greater distance between host and native countries enhances cultural retention. Overall, our findings contribute to a better understanding of the multifaceted factors shaping culture retention among immigrants, offering valuable insights to understand cultural diversity within host countries.

\begin{figure*}[ht!]
\centering
\includegraphics[width=\linewidth]
{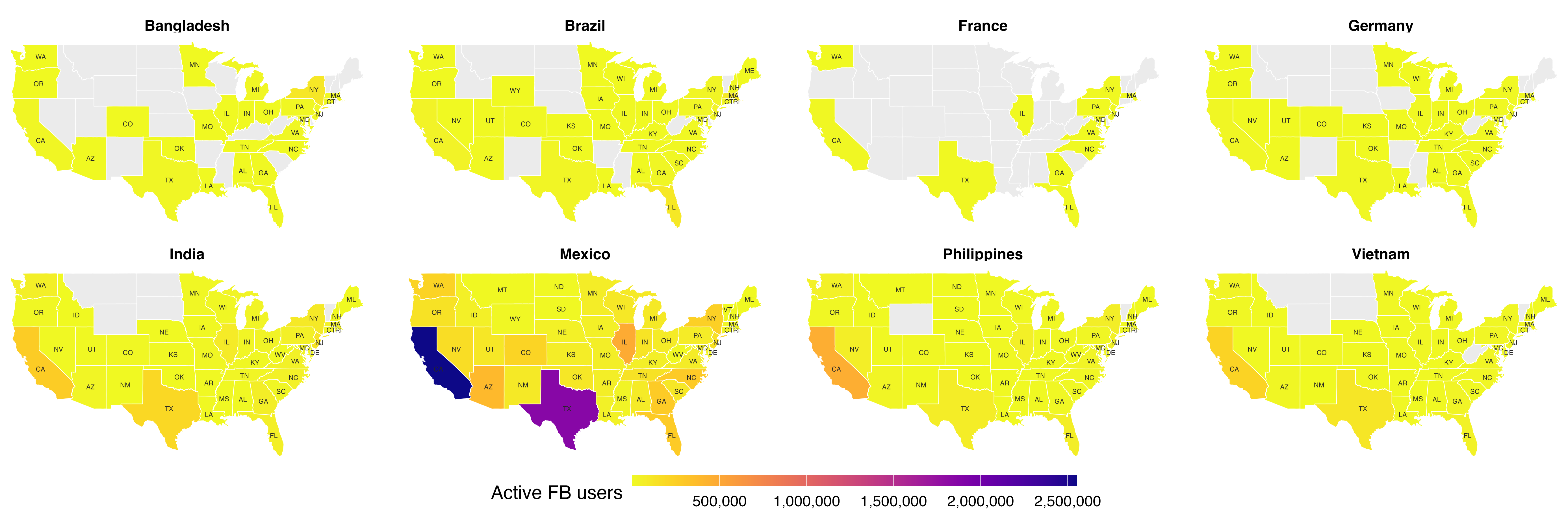} 
\caption{Spatial distribution of \textit{Active Facebook Users} across U.S. states among 8 diaspora groups, i.e., expats, according to Facebook categorization. Grey states indicate data unavailability or sparsity. }
\label{fig:Geographic distribution}
\end{figure*}

\section{Melting Pot or Mosaic? }
The metaphor of the melting pot emphasizes assimilation: immigrants are expected to embrace the host nation’s norms and values, often linked to higher well-being and social integration. A plethora of studies probed the acculturation process of immigrants in the host country and the consequences of the same. For instance, the literature suggests a positive association between cultural assimilation (of immigrants with the host culture) and immigrants’ life satisfaction or subjective well-being, i.e., a strong host identity is associated with high life satisfaction, but this is not the case for native identity \cite{angelini2015life}. A weak sense of native identity is associated with higher levels of life satisfaction. Interestingly, these results remain consistent even after controlling for individual characteristics like employment status and income, and notably, “potential benefits of cultural assimilation go beyond the time dimension of the integration process … its association with life satisfaction is stronger for established immigrants than for recent ones” \cite[p.~841]{angelini2015life}. Other studies also pointed out that challenges in cultural assimilation or adaptation are often associated with long-term mental health problems like trauma and distress \cite{bhugra2005cultural}. 

Conversely, the mosaic hypothesis highlights cultural pluralism: immigrants preserve distinct practices and traditions that coexist with those of the host society. Extant literature has highlighted that immigrants retain their native culture in terms of leisure practices \cite{heinonen2005leisure}, food and dietary preferences \cite{lillekroken2024food}. This pluralism can enrich the host country by fostering mutual understanding. For instance, the US president has been wishing and celebrating `Happy Diwali', a festival of light primarily celebrated in India \cite{whitehouse2022}. This greeting conveys respect and cultural recognition of the Indian American community settled in the USA. These practices lend support to the mosaic view. 

Overall, this debate connects to broader theories of acculturation, such as the seminal work of \citet{berry1997immigration}, which distinguishes between \textit{assimilation }(when individuals adopt the host culture over their own), \textit{separation } (when they reject the host culture to preserve their origin culture), \textit{integration }(when they embrace aspects of both cultures), and \textit{marginalization }(when they disengage from both cultures). The Melting Pot Hypothesis closely aligns with Berry’s assimilation, whereas the Mosaic Hypothesis emphasizes pluralism and multiculturalism traditions, i.e., Berry’s integration categorization, emphasizing the coexistence of multiple cultural identities within a host society. However, it is quite possible that an immigrant may reject the host culture to preserve their native culture, i.e., separation. Broadly, while the melting pot assumes eventual convergence into a dominant culture, the mosaic points out that immigrants may add to the cultural diversity of the host nation. 

Notably, these acculturation theories were developed primarily in offline contexts. We argue that cultural retention on online platforms may manifest differently. As noted, immigrants may continue their interests in home-country music, cuisine, or media, even as they participate in offline host-country cultural practices and social activities. By examining how origin-country characteristics and host–native distances shape such online expressions of culture, our study enriches the theoretical discourse of acculturation by providing one of the first large-scale empirical studies on online cultural retention.

\section{Migration Studies using Digital Data}
In recent times, extant literature has used digital data to study the cultural dynamics of immigrants. Specifically, Facebook data was considered as a proxy for immigrants' cultural interests \cite{stewart2019rock, dubois2018studying, vieira2020using}. This stream of studies mostly examined the cultural assimilation of migrants, i.e., the melting pot hypothesis. For instance, \citet{stewart2019rock} argued that the alignment of musical tastes with host-country preferences indicates higher assimilation and found that Mexican and Indian immigrants in the U.S. exhibited relatively higher cultural convergence than immigrants from East Asia. Similarly, \citet{dubois2018studying} highlighted the role of language and origin-country characteristics in shaping assimilation patterns among Arabic-speaking migrants in Germany.

Beyond cultural aspects, extant literature has primarily focused on opinion mining and digital census applications. Twitter-based studies, presently $\mathbb{X}$ platform, reveal polarized sentiments toward immigrants: some of the prior studies emphasized a humanitarian perspective, such as struggles of migrants \cite{siapera2018refugees, khatua2021struggle,khatua2022rites,khatua2022endorsement,khatua2021analyzing}, while other studies portrayed migrants as cultural or security threats \cite{kreis2017refugeesnotwelcome,khatua2023we}. Another stream of research considered online data for performing a digital census, such as estimating migration stocks and flows using Facebook advertising data \cite{zagheni2017leveraging,rampazzo2021framework}
 or geo-tagged Twitter traces \cite{zagheni2014inferring, kim2020digital}. These studies confirm the viability of digital data as an alternative to surveys and censuses. A handful of qualitative studies elucidate that online platforms also provide emotional support and continuity for immigrants, offering ways to sustain connections with families and home cultures \cite{francisco2015internet,lingel2014city,kaufmann2018navigating,khatua2025policies}.

\textbf{Research Gap:} In short, extant literature, using digital data, has primarily emphasized assimilation, reinforcing the melting pot perspective. Other studies also point out that social media enables migrants to maintain cultural ties, implicitly supporting the mosaic hypothesis. However, to the best of our knowledge, \textit{no study has systematically examined the conditions under which online cultural retention emerges}, or the role of origin-country characteristics and host–native distances in shaping it. Our study addresses this gap by analyzing Facebook advertising data on the cultural interests of immigrants from 8 countries in the USA. We move beyond cultural assimilation to identify factors that foster or constrain cultural retention, to \textit{investigate the mosaic hypothesis in a digital context}.

\begin{table*}[ht!]
\captionsetup{font=scriptsize} 
 \scriptsize
    \centering
    \begin{tabular}{r|p{14cm}}
        \toprule
        \textbf{Country} & \textbf{Country-specific Cultural Interests} \\
        \midrule
        Bangladesh \ctry{BD} & Bangladeshi rock, Cinema of Bangladesh, Bangladesh national cricket team, Bengali cuisine, RTV (Bangladesh), Bangladesh Premier League, Bangladesh Football Federation, Nemesis (Bangladeshi band), Black (Bangladeshi band) \\
        \midrule
        Brazil \ctry{BR}& Music of Brazil, Brazilian rock, Bossa nova, Brazilian cuisine, Texas de Brazil, Churrascaria, Brazil national football team, Brazil national under-20 football team, Brazilian jiu-jitsu, Campeonato Brasileiro Série A, Campeonato Brasileiro Série B \\
        \midrule

        France \ctry{FR}& Culture of France, French cuisine, French wine, Baguette, French hip hop, French pop music, Hellfest (French music festival), France Télévisions, Cinema of France, France national football team, France Football, France national rugby union team, Coupe de France \\
        \midrule
        Germany \ctry{DE} & Cinema of Germany, German television comedy, The Voice of Germany, German hip hop, German cuisine, Sport1 (Germany), Germany national football team, Germany national under-21 football team, German football league system, Bundesliga, German Grand Prix \\
        \midrule

        India \ctry{IN} & Indian rock, Indian pop, Indian folk music, Indian classical music, Music of India, Cinema of India, Indian cuisine, South Indian cuisine, North Indian cuisine, Indian fast food, Indian Premier League, Indian Super League, Indian Cricket League, Indian Cricket \\
        \midrule
        Mexico \ctry{MX}& Music of Mexico, Cinema of Mexico, Regional styles of Mexican music, Mexican pop music, Beer in Mexico, Mexican cuisine, Mexico national football team, Liga MX, Mexican League, Mexican Pacific League, Mexican Football Federation \\
        \midrule
        Philippines \ctry{PH}& Music of the Philippines, Sugarfree (Filipino band), Pinoy pop, TV5 (Philippines), Filipino Style Recipe, Philippine cuisine, Mang Inasal Philippines, Filipino martial arts, University Athletic Association of the Philippines, Philippines men's national basketball team \\
        \midrule
        Vietnam \ctry{VN}& Vietnamese cuisine, Music of Vietnam, Vietnamese literature, The Voice of Vietnam, MTV (Vietnam), Vietnam national football team, Tet, Vietnam national under-23 football team, Vietnam's Next Top Model \\
        
        \bottomrule
    \end{tabular}
    \caption{Country-specific interests from \textbf{Facebook Advertising API identified for capturing Cultural Retention of immigrants}. }
\label{tab:Country_specific_interests}
\end{table*}

\section{Methodology}
\textbf{Country Selection:} We selected a total of 8 countries from Asia, Southeast Asia, Latin America, and Europe, 2 from each region, based on the higher presence of immigrants in the USA. We had to consider the availability of data. For instance, the highest numbers of immigrants in the United States are from Mexico, China, and India\footnote{https://worldpopulationreview.com/country-rankings/us-immigration-by-country}. However, due to the scarcity of Facebook data in China, we could not consider it in our analysis. {Figure \ref{fig:Geographic distribution}} shows the state-wise distribution of immigrants, and given the data sparsity even for the 8 countries included, adding more countries would contribute little additional value. Thus, while the list is not exhaustive, it is appropriate to address our research questions.

\textbf{Facebook Advertising API:} Facebook users leave traces of their preferences across diverse interests such as \textit{music, food, and recreation} through interactions like watching videos or commenting on posts. For a specific interest, Facebook provides advertisers an estimated audience size, i.e., the total number of active Facebook users for specific demographics (e.g., age, gender, or education). While it's not explicitly outlined how these interests are generated or maintained by Facebook, previous research indicates that these interests are dynamic and subject to change over time \cite{,stewart2019rock, dubois2018studying}. 
While some interests are shared across most countries, others are highly country-specific \cite{vieira2020using, vieira2022interplay}. Overall, Facebook allows the extraction of data for a diverse range of interests tailored to different countries \cite{obradovich2022expanding}.

\begin{figure}[ht!]
\centering
\includegraphics[width=1\columnwidth]{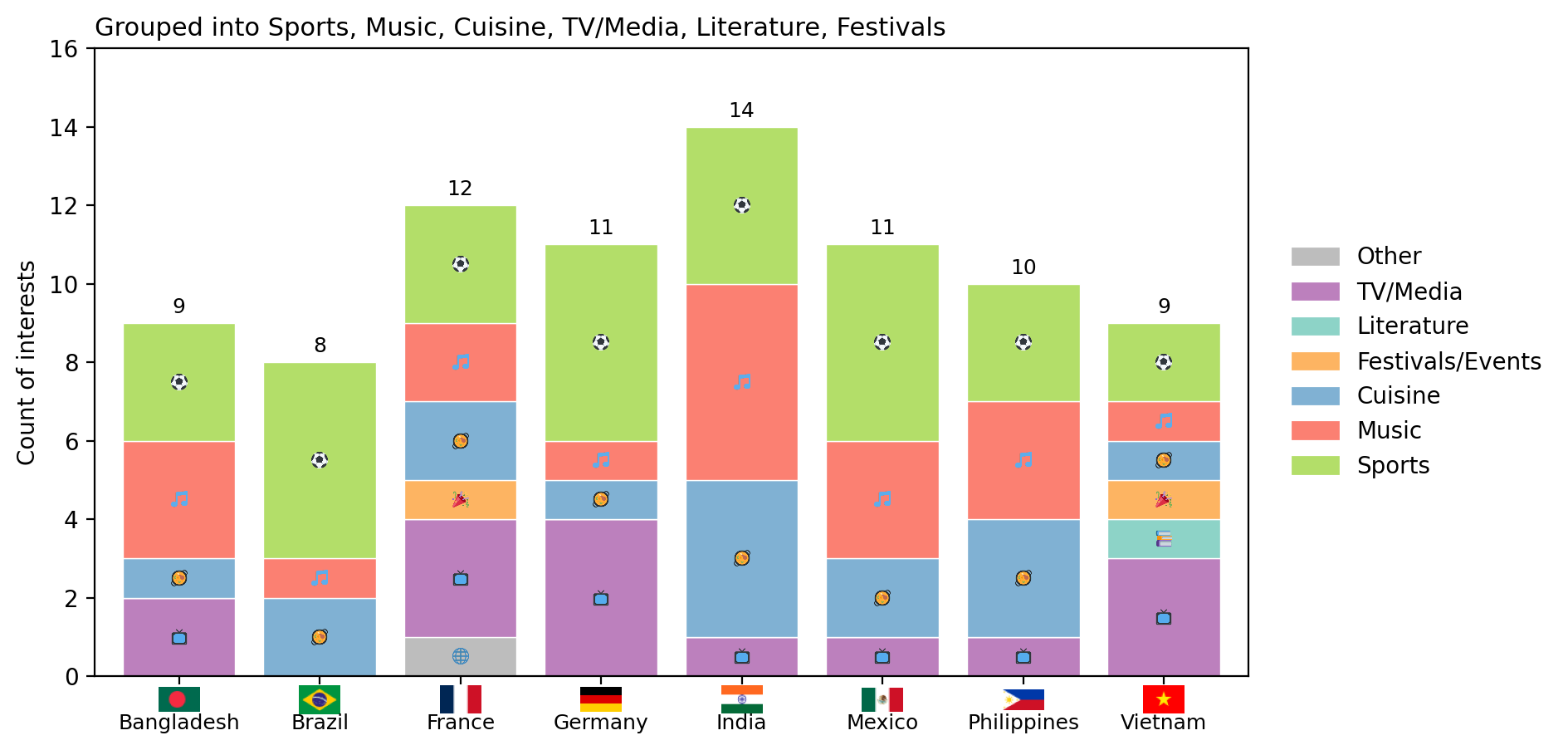} 
\caption{Bar chart showing the distribution of interests across different countries, grouped by categories. The chart reports the count of interests for each category in 8 countries. Icons within the bars provide visual indicators of specific interests within each group.} 
\label{fig:grouped Interests}
\end{figure}

Accordingly, our focus primarily revolves around country-specific interests; specifically, we consider a wide range of \textit{country-specific cultural interests} - as listed in Table \ref{tab:Country_specific_interests}. The patterns of interests significantly vary across countries \cite{vieira2020using,obradovich2022expanding}; for instance, India boasts a diverse range of cuisine-related interests based on regional distinctions, whereas Germany lacks this diversity. {Figure \ref{fig:grouped Interests}} graphically reports these variations across countries. We do agree that it may not be feasible to capture an entire country's cultural orientation through a few interests, but this limitation lies beyond our control and is determined by Facebook's algorithms, which aim to reach and engage the maximum number of users. Thus, despite these limitations, it is also worth acknowledging that Facebook's enormous revenue and profit stem from selling this data for marketing purposes. This implicitly affirms the validity and reliability of these interests.

Additionally, Facebook has the option to specify the geographical location from which we wish to retrieve audience estimates for a specific interest. Also, the Facebook API allows the extraction of information about immigrants (expats). For example, we can extract audience sizes who previously resided in one country (e.g., Mexico) but currently live elsewhere (e.g., the USA). This option allows us to effectively estimate the number of immigrants (i.e., audience size) from India or Mexico now living in the USA \cite{khatua2023host,dubois2018studying,stewart2019rock}. When estimating audience size within their home country, we set the location/country accordingly. However, for the host country, in this study, we set it to the USA.

\subsection{Operationalization of Variables} 

\textbf{Cultural Retention}: We measure \textit{cultural retention of immigrants} as their continued interest in their home country's traditions - specifically in the context of culinary choices \cite{lillekroken2024food, vieira2022interplay}, musical \cite{stewart2019rock, ripani2023music}, and recreational preferences \cite{heinonen2005leisure,dubois2018studying}. The proliferation of various online platforms allows an immigrant to maintain her connection to her roots \cite{barnett2017predicting,croucher2015longitudinal,gunsoy2020cultural}. Online platforms enable immigrants to watch movies or TV shows from their native countries and exchange opinions (via online platforms) with family and friends residing in their respective native countries. For instance, the cultural retention of French immigrants can be captured through their digital traces (i.e., watching videos or commenting on posts) related to French cultures (i.e., France-specific interests listed in Table \ref{tab:Country_specific_interests}). If more French immigrants are following or interacting with French-culture-specific content, this would indicate higher cultural retention by them.

\begin{table*}[ht!]
\scriptsize
\centering

\begin{tabular}{@{}lccccrcr@{}}
\toprule
Variables & Mean & S.D. & Minimum & Country with Lowest Score  & Maximum & Country with Highest Score \\ 
\midrule
Ln(GDP) & 14.08 & 0.983 & 12.91 & \ctry{PH} & 15.222 & \ctry{DE} \\
Ln(GDP/capita) & 9.005 & 1.183 & 7.788 & \ctry{IN}  & 10.794 & \ctry{DE} \\
Political Rights Score & 26.375 & 11.856 & 4.00 & \ctry{VN} & 39.00 & \ctry{DE} \\
Civil Liberties Score & 35.75 & 12.948 & 15.00 & \ctry{VN} & 54.00 & \ctry{DE} \\
Linguistic Diversity Index & 0.393 & 0.311 & 0.100 & \ctry{BR} & 0.910 & \ctry{IN}  \\
Int’l Tourists Arrival/Population & 0.599 & 1.099 & 0.002 & \ctry{BD} & 3.233 & \ctry{FR} \\
Ln(Immigrants in the USA) & 12.33 & 1.233 & 10.69 & \ctry{FR} & 14.208 & \ctry{MX} \\
Human Development Index & 0.769 & 0.109 & 0.644 & \ctry{IN} & 0.95 & \ctry{DE} \\
Ln(Geographic Distance) & 9.079 & 0.579 & 7.811 & \ctry{MX} & 9.526 & \ctry{VN} \\
Ln(GDP Difference) & 16.974 & 0.062 & 16.877 & \ctry{DE} & 17.036 & \ctry{PH} \\
Ln(GDP/capita Difference) & 10.963 & 0.386 & 10.226 & \ctry{DE} & 11.211 & \ctry{IN}  \\
Ln(Power Distance Difference)  & 4.198 & 0.235 & 3.829 & \ctry{PH} & 4.654 & \ctry{DE} \\
Ln(Individualism Difference) & 4.786 & 0.235 & 4.394 & \ctry{DE} & 5.043 & \ctry{BD} \\
Ln(MAS Difference) & 4.666 & 0.100 & 4.533 & \ctry{MX} & 4.804 & \ctry{VN}\\
Ln(Uncertainty Avoidance Difference) & 4.424 & 0.244 & 4.094 & \ctry{FR} & 4.754 & \ctry{VN}\\
Ln(Long Term Orientation Difference) & 4.659 & 0.123 & 4.500 & \ctry{FR}  & 4.844 & \ctry{MX} \\
Ln(Indulgence Difference) & 4.784 & 0.231 & 4.263 & \ctry{MX} & 4.997 & \ctry{BD} \\
\bottomrule
\end{tabular}
\caption{Data Overview of Variables}
\label{Tab:data_snapshot}
\end{table*}

However, absolute values (i.e., the estimated audience size of these French immigrants) would not reveal much insight. Hence, we need to normalize the same with a base value, i.e., French Facebook users in the USA. Next, we also need to consider the culture-related interests of the French native population staying in their home/native country. To estimate audience size, we consider MAU (monthly active users). Facebook API returns two MAU values: lower bound and higher bound. We have taken an average of these two values as audience size \cite{khatua2023host}. However, audience size is a function of Facebook users. Hence, a country with higher Facebook penetration will have a higher audience size. To tackle these distortions, we normalize the estimated audience size with total Facebook users. Also, Facebook API may return an audience value less than 1000 - indicating \textit{sparsity}, i.e., Facebook doesn't have a sufficient number of users for that specific category. To control sparsity, we exclude all data points (equal to or less than 1000) from our final analysis.

\begin{figure}[ht!]

\centering
\includegraphics[width=1\columnwidth, height=5.5 cm]{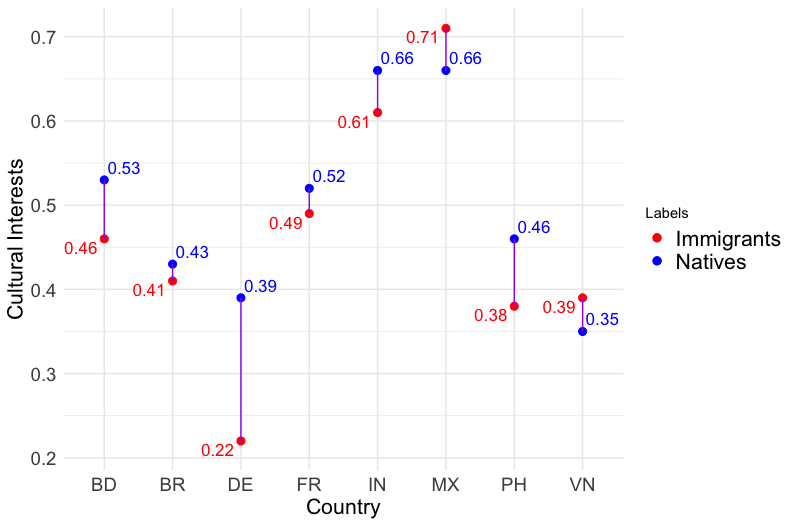} 
\caption{\textcolor{red}{Red dots} report the median values of \textit{$CI(Imm)_{\text{Ctry}}$},i.e., cultural interests across various demographic categories of immigrants residing in the US. 
Conversely, \textcolor{blue}{blue dots} report the median values of \textit{$CI(Nat)_{\text{Ctry}}$},i.e., similar cultural interests of natives - staying in their respective home countries. This figure indicates wide cross-country variations, e.g., higher values for DE and lower values for BR and VN.} 
\label{fig: median Immigrants vs Natives}
\end{figure}

Following \citet{dubois2018studying} and \citet{stewart2019rock}, we estimate \textit{Cultural Retention (CR)} for a group of immigrants with respect to natives residing in their home country. Say, we are estimating the cultural orientation of German immigrants in the USA, i.e., our host country ($\text{Host}_{\text{Ctry}}$). To estimate \textit{Cultural Interests of Immigrants from a specific native country, i.e., $CI(Imm)_{\text{Ctry}}$}, we need to estimate the proportion of German immigrants interested in \textit{distinct} Germany or Country-specific Cultural Interests $(\textit{Ctry}_{\textit{CI}})$. 

\begin{equation}
    CI(Imm)_{\text{Ctry}} = \frac{\# \text{ of immigrants interested in Ctry}_{\text{CI}}}{\# \text{ of immigrants in the $\text{Host}_{\text{Ctry}}$}}
\end{equation}

Next, to estimate \textit{Cultural Interests of Natives of the same country, i.e., $CI(Nat)_{\text{Ctry}}$}, we need to estimate the proportion of natives (or Germans residing in Germany) displaying preferences for the \textit{same set of distinct} German or Country-specific Cultural Interests $(\textit{Ctry}_{\textit{CI}})$. 

\begin{equation}
    CI(Nat)_{\text{Ctry}} = \frac{\# \text{ of natives interested in Ctry}_{\text{CI}}}{\# \text{ of natives in the $\text{Home}_{\text{Ctry}}$}}
\end{equation}

Conceptually, similar values of \textit{$CI(Imm)_{\text{Ctry}}$} and \textit{$CI(Nat)_{\text{Ctry}}$} will indicate no significant differences between immigrants and natives, i.e, broadly similar cultural orientations irrespective of where they are residing. However, {Figure \ref{fig: median Immigrants vs Natives}} indicates variations between natives and immigrants. 
Subsequently, we also explore the state-wise distribution of cultural interests of immigrants, i.e., \textit{$CI(Imm)_{\text{Ctry}}$}, and report it in {Figure \ref{fig:Own-culture engagement}}. With the exception of a few countries, such as Mexico or India, and a few states, such as Texas, data scarcity remains a major concern. This limitation prevents us from conducting a granular sub-national analysis and restricts our analysis to the country level.  
Following \citet{dubois2018studying} and \citet{stewart2019rock}, we operationalize \textit{Cultural Retention (CR)} as follows: 

\begin{equation}
\text{CR$_1$} = \frac{\text{$CI(Imm)_{\text{Ctry}}$}}{\text{$CI(Nat)_{\text{Ctry}}$}}
\end{equation}

\begin{figure*}[ht!]
\centering
\includegraphics[width=\linewidth]{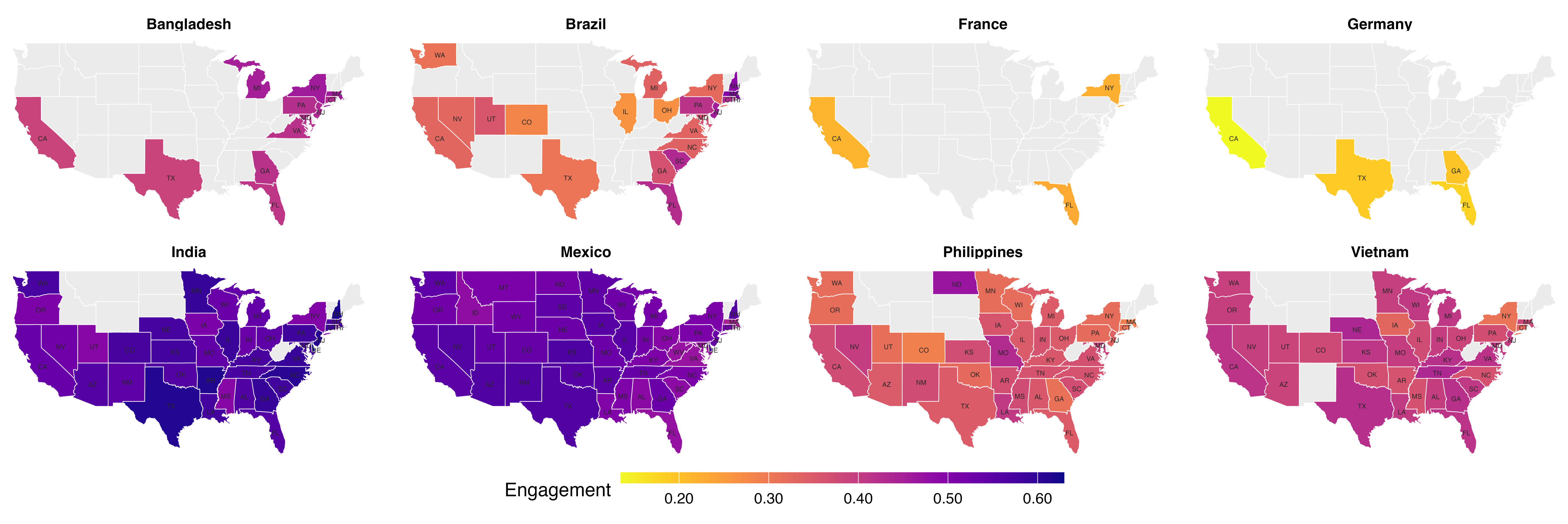} 
\caption{State-level distribution of \textit{own-culture engagement}, shown as the spatial pattern of \textcolor{red}{red dots} in {Figure \ref{fig: median Immigrants vs Natives}}, for 8 diaspora groups. Grey states denote data sparsity, which is higher compared to {Figure \ref{fig:Geographic distribution}} because {Figure \ref{fig:Geographic distribution}} reports the presence of expats (or absolute values of audience size), whereas this figure reports state-wise median values of \textit{$CI(Imm)_{\text{Ctry}}$}.}
\label{fig:Own-culture engagement}
\end{figure*}

\begin{table*}[ht!]
\begin{center}
\captionsetup{font=footnotesize} 
\scriptsize
\begin{tabular}{l c c c c c c c c c}
\toprule
 & \multicolumn{1}{c}{Model 1} & \multicolumn{1}{c}{Model 2} & \multicolumn{1}{c}{Model 3} & \multicolumn{1}{c}{Model 4} & \multicolumn{1}{c}{Model 5} & \multicolumn{1}{c}{Model 6} & \multicolumn{1}{c}{Model 7} & \multicolumn{1}{c}{Model 8} & \multicolumn{1}{c}{Model 9} \\
\midrule
Intercept         & $\textcolor{blue}{0.97^{***}}$ & $\textcolor{blue}{1.41^{***}}$ & $\textcolor{blue}{1.11^{***}}$ & $\textcolor{blue}{1.19^{***}}$  & $\textcolor{blue}{1.34^{***}}$  & $\textcolor{blue}{1.05^{***}}$  & $\textcolor{blue}{0.97}$       & $-0.17$      & $\textcolor{blue}{1.12^{***}}$ \\
Age: 35 to 55     & $0.03$       & $0.04$       & $0.04$       & $0.04$        & $0.05$        & $0.03$        & $0.02$       & $0.04$       & $0.04$       \\
Age: 56 and above & $0.05$       & $0.05$       & $0.05$       & $0.07$        & $0.08^{*}$    & $0.05$        & $0.02$       & $0.05$       & $0.05$       \\
Male Dummy        & $-0.08^{*}$  & $-0.08^{*}$  & $-0.08^{*}$  & $-0.09^{**}$  & $-0.10^{**}$  & $-0.07^{*}$   & $-0.07^{*}$  & $-0.09^{**}$ & $-0.08^{*}$  \\
Edu: Advanced     & $-0.09^{*}$  & $-0.09^{*}$  & $-0.09^{*}$  & $-0.07$       & $-0.06$       & $-0.09^{*}$   & $-0.10^{**}$ & $-0.08^{*}$  & $-0.09^{*}$  \\
Edu: Basic        & $-0.04$      & $-0.04$      & $-0.04$      & $-0.04$       & $-0.04$       & $-0.04$       & $-0.06$      & $-0.05$      & $-0.04$      \\
Economic Status: iOS users & $0.02$       & $0.02$       & $0.02$       & $0.02$        & $0.02$        & $0.02$        & $0.02$       & $0.03$       & $0.02$       \\
Ln(GDP)           &              & $-0.03$      &              &              &              &              &              &              &              \\
Ln(GDP/capita)    &              &              & $-0.02$      &              &              &              &              &              &              \\
Political Rights Score &          &              &              & $\textcolor{blue}{-0.01^{***}}$ &              &              &              &              &              \\
Civil Liberties Score  &          &              &              &              & $\textcolor{blue}{-0.01^{***}}$ &              &              &              &              \\
Linguistic Diversity Index &      &              &              &              &              & $\textcolor{blue}{-0.18^{***}}$ &              &              &              \\
Int’l Tourists Arrival/Population & &            &              &              &              &              & $0.03$       &              &              \\
Ln(Immigrants in the USA) &       &              &              &              &              &              &              & $\textcolor{blue}{0.09^{***}}$ &              \\
Human Development Index &         &              &              &              &              &              &              &              & $-0.21$       \\
\midrule
$R^2$             & $0.05$       & $0.07$       & $0.06$       & $0.23$        & $0.28$        & $0.11$        & $0.16$       & $0.16$       & $0.06$       \\
Adj. $R^2$        & $0.03$       & $0.04$       & $0.02$       & $0.20$        & $0.25$        & $0.08$        & $0.13$       & $0.13$       & $0.03$       \\
Max. VIF value    & $1.01$       & $1.02$       & $1.04$       & $1.02$        & $1.03$        & $1.03$        & $1.01$       & $1.02$       & $1.05$       \\
Num. of Obs.         & $214$        & $214$        & $214$        & $214$         & $214$         & $214$         & $214$        & $214$        & $214$        \\
\bottomrule
\end{tabular}
\caption{Reports the OLS regression analysis regarding the \textbf{Effects of Country Characteristics on Culture Retention of Immigrants}. Here, base categories for demographic variables for all regression models are Age: 18-34, Gender: Female, Edu: Not disclosed, and OS: Android Users. \textbf{Dependent Variables: \text{CR$_1$}}. Significance level: $^{***}p<0.001$; $^{**}p<0.01$; $^{*}p<0.05$. \textbf{Robustness tests}: \textcolor{blue}{Bonferroni Correction} reconfirms the OLS-based regression analysis at $p<0.05$ for the \textcolor{blue}{variables color-coded in blue}. }
\label{tab:CR1_CountryCharacteristics}
\end{center}
\end{table*}

For robustness, we also consider CR as follows: 

\begin{equation}
\text{CR$_2$} = Ln\frac{\text{$CI(Imm)_{\text{Ctry}}$}}{\text{$CI(Nat)_{\text{Ctry}}$}}
\end{equation}

\begin{table*}[ht!]
\scriptsize
\captionsetup{font=footnotesize} 

\begin{center}
\begin{tabular}{l c c c c c c  c c c}
\toprule
 & \multicolumn{1}{c}{Model 1} & \multicolumn{1}{c}{Model 2} & \multicolumn{1}{c}{Model 3} & \multicolumn{1}{c}{Model 4} & \multicolumn{1}{c}{Model 5} & \multicolumn{1}{c}{Model 6} & \multicolumn{1}{c}{Model 7} & \multicolumn{1}{c}{Model 8} & \multicolumn{1}{c}{Model 9} \\
\midrule
Intercept         & $\textcolor{blue}{0.67^{***}}$ & $\textcolor{blue}{0.83^{***}}$ & $\textcolor{blue}{0.75^{***}}$ & $\textcolor{blue}{0.77^{***}}$  & $\textcolor{blue}{0.85^{***}}$  & $\textcolor{blue}{0.71^{***}}$  & $0.67^{***}$ & $0.03$       & $\textcolor{blue}{0.77^{***}}$ \\
Age: 35 to 55 & $0.02$       & $0.02$       & $0.02$       & $0.02$        & $0.02$        & $0.01$        & $0.01$       & $0.02$       & $0.02$       \\
Age: 56 and above   & $0.02$       & $0.02$       & $0.02$       & $0.03$        & $0.03$        & $0.02$        & $0.02$       & $0.02$       & $0.02$       \\
Male Dummy          & $-0.04^{*}$  & $-0.04^{*}$  & $-0.04^{*}$  & $-0.05^{**}$  & $-0.05^{**}$  & $-0.04^{*}$   & $-0.04^{*}$  & $-0.05^{**}$ & $-0.04^{*}$  \\
Edu: Advanced   & $-0.04$      & $-0.04$      & $-0.04$      & $-0.03$       & $-0.03$       & $-0.04^{*}$   & $-0.04^{*}$  & $-0.04$      & $-0.04$      \\
Edu: Basic   & $-0.02$      & $-0.02$      & $-0.02$      & $-0.02$       & $-0.02$       & $-0.02$       & $-0.02$      & $-0.02$      & $-0.02$      \\
Economic Status: iOS users          & $0.01$       & $0.01$       & $0.01$       & $0.01$        & $0.01$        & $0.01$        & $0.01$       & $0.01$       & $0.01$       \\
Ln(GDP)             &            & $-0.01$      &            &             &             &             &            &            &            \\
Ln(GDP/capita)           &            &            & $-0.01$      &             &             &             &            &            &            \\
Political Rights Score          &            &            &            & $\textcolor{blue}{-0.00^{***}}$ &             &             &            &            &            \\
Civil Liberties Score               &            &            &            &             & $\textcolor{blue}{-0.01^{***}}$ &             &            &            &            \\
Linguistic Diversity Index                &            &            &            &             &             & $\textcolor{blue}{-0.09^{***}}$ &            &            &            \\
Int’l Tourists Arrival/Population        &            &            &            &             &             &             & $0.01$       &            &            \\
Ln(Immigrants in the USA)           &            &            &            &             &             &             &            & $\textcolor{blue}{0.05^{***}}$ &            \\
Human Development Index                 &            &            &            &             &             &             &            &            & $-0.14$      \\
\midrule
$R^2$               & $0.05$       & $0.06$       & $0.05$       & $0.19$        & $0.25$        & $0.10$        & $0.05$       & $0.18$       & $0.06$       \\
Adj. $R^2$          & $0.02$       & $0.02$       & $0.02$       & $0.16$        & $0.22$        & $0.07$        & $0.02$       & $0.15$       & $0.02$       \\
Max. VIF value         & $1.00$        & $1.02$         & $1.04$       & $1.02$         & $1.03$       & $1.00$  & $1.02$  & $1.02$ & $1.05$  \\
Num. obs.           & $214$        & $214$        & $214$        & $214$         & $214$         & $214$         & $214$        & $214$        & $214$        \\
\bottomrule
\multicolumn{10}{l}{\scriptsize{$^{***}p<0.001$; $^{**}p<0.01$; $^{*}p<0.05$}}
\end{tabular}
\caption{\textbf{Robustness Analysis: Effects of Country Characteristics on Culture Retention of Immigrants}. Here, base categories for demographic variables for all regression models are Age: 18-34, Gender: Female, Edu: Not disclosed, and OS: Android Users. Dependent Variables: $\text{CR}_2$. Significance level: $^{***}p<0.001$; $^{**}p<0.01$; $^{*}p<0.05$. \textcolor{blue}{Bonferroni Correction} reconfirms the OLS-based regression analysis at $p<0.05$ for the \textcolor{blue}{variables color-coded in blue}.}
\label{tab:CR2_CountryCharacteristics}
\end{center}
\end{table*}

\textbf{Native Country Characteristics}: We have considered country-level factors that are likely to shape migrants' values, norms, and behavioral patterns, thereby influencing their cultural retention.

\textit{GDP and GDP/capita}: The economic status or affordability of an immigrant might impact their ability to embrace the host culture. Hence, we consider the Gross Domestic Product (GDP) of a country (in current USD) and GDP per capita (in current USD) to capture this aspect. We extracted data for these variables from the World Bank\footnote{https://data.worldbank.org/indicator}. To address significant disparities among countries, we logarithmically transformed these variables for our analysis.

\textit{Human Development Index}: Unlike the unidimensional GDP/capita, the Human Development Index (HDI) is a composite index that considers three basic dimensions of human development: Health, Education, and Standard of living. Thus, HDI effectively captures the overall well-being and quality of life in the native country. Therefore, in our analysis, we also consider HDI\footnote{https://hdr.undp.org/data-center/country-insights\#/ranks}.

\textit{Political Rights and Civil Liberties}: Socio-political freedom in the native country can influence behavior and attitude towards diversity or plurality in opinions, which can, in turn, negatively impact the propensity of cultural retention, \textit{refuting the mosaic hypothesis}. Hence, we consider these two aspects in our analysis and extract the data from the Freedom House website\footnote{https://freedomhouse.org/countries/freedom-world/scores}. Higher scores indicate greater freedom.

\textit{Linguistic Diversity}: Similar to socio-political freedom, linguistic diversity, i.e., the number of languages spoken within a country, varies significantly across countries. Some nations have only one language, while others have multiple languages or at least different dialects spoken in various regions. We assume exposure to diverse cultures/languages would negatively impact cultural retention, \textit{discarding the mosaic hypothesis}. We consider the Linguistic Diversity Index\footnote{https://worldpopulationreview.com/country-rankings/linguistic-diversity-index-by-country} for our analysis.

\textit{Exposure to international tourism}: The opportunity to interact with international tourists in the native country exposes locals (i.e., potential immigrants) to different cultures, fostering greater cultural understanding. Thus, immigrants from countries known for international tourism may have a lower propensity to retain native culture, \textit{contradicting the mosaic hypothesis}. We captured this aspect by dividing international tourism (i.e., number of arrivals) by population. To nullify the COVID-related negative impact on tourism, we used 2019 data points from the World Bank\footnote{https://data.worldbank.org/indicator}. 

\textit{Exposure to Compatriots}: Conversely, the opportunity to interact with compatriots or co-nationals in the host country allows an immigrant to maintain their own cultural identities. These immigrants may gather or come together during festivals or cultural occasions, \textit{confirming the mosaic hypothesis}. We capture this effect by incorporating country-wise data points of USA immigrants\footnote{https://worldpopulationreview.com/country-rankings/us-immigration-by-country}. To control heteroscedasticity, we performed a log transformation.

\textbf{Different Dimensions of Distance}: For our RQ2, we have considered three dimensions of distance as follows: geographic distance, economic distance, and cultural distance \cite{coimbra2024value}.

\textit{Geographic Distance:} 

The greater the geographic distance between two locations, the more challenging it can be to travel or communicate. 

Consequently, less frequent travel to the native country can encourage immigrants to adopt the host culture. Considering the wide disparities in this variable, we have performed log transformations of geographic distances \cite{CEPII:2011-25}.

\textit{Economic Distance:} Previously, we considered the economic status of native countries, but  

it is worth noting that the disparity between two countries can be a \textit{better predictor of cultural retention or the mosaic hypothesis}. For instance, it would often be challenging for an immigrant from a less economically developed country to afford or participate in the cultural activities of a more affluent country like the USA. Therefore, we considered the difference or gap in GDP and GDP/capita between the USA and the respective native countries and performed a log transformation on these variables for our analysis.

\textit{Cultural Distances:} Following \citet[p.~147]{tadesse2010cultural}, we conceptualize cultural distance “as a measure of the degree to which shared norms and values in one country differ from those in another country.” We consider cultural distances across \textit{Hofstede’s 6 Cultural Dimensions} \cite{hofstede1984hofstede, hofstede1988confucius,taras2010examining}

However, is Hofstede’s framework, developed in the 1960s and 1970s, still relevant today? \citet{beugelsdijk2015scores} argue that while cultures have changed slightly, the relative differences between countries remain largely the same, making the framework a reliable measure of cultural distance. 
We operationalize this variable as follows: Hofstede’s Individualism dimension scores for the USA, Brazil, and Vietnam are 60, 36, and 30, respectively. Therefore, we calculate the \textit{Individualism Difference} as \textit{Ln(60-36)} for Brazil and \textit{Ln(60-30)} for Vietnam. However, it is worth noting that some countries may have higher scores than the USA on certain cultural dimensions. To avoid negative values, we take the absolute value before performing logarithmic transformations.

\begin{table*}[ht!]
\begin{center}
\captionsetup{font=footnotesize}
\scriptsize
\begin{tabular}{l c c c c c c c}
\toprule
 & \multicolumn{1}{c}{Model 1} & \multicolumn{1}{c}{Model 2} & \multicolumn{1}{c}{Model 3} & \multicolumn{1}{c}{Model 4} & \multicolumn{1}{c}{Model 5} & \multicolumn{1}{c}{Model 6} & \multicolumn{1}{c}{Model 7} \\
\midrule
Intercept         & $\textcolor{blue}{1.44^{***}}$ & $-2.76^{**}$ & $-10.44$    & $-6.86^{**}$  & $-0.62$     & $\textcolor{blue}{-8.29^{***}}$ & $\textcolor{blue}{0.79^{***}}$ \\
Age: 35 to 55     & $0.03$       & $0.05$       & $0.04$      & $0.05$        & $0.04$      & $0.06^{*}$    & $0.05$ \\
Age: 56 and above & $0.04$       & $0.08$       & $0.06$      & $0.06^{*}$    & $0.05$      & $0.06^{*}$    & $0.06^{*}$ \\
Male Dummy        & $-0.07^{*}$  & $-0.10^{**}$ & $-0.08^{*}$ & $\textcolor{blue}{-0.10^{***}}$ & $-0.08^{*}$ & $\textcolor{blue}{-0.10^{***}}$ & $\textcolor{blue}{-0.10^{***}}$ \\
Edu: Advanced     & $-0.09^{*}$  & $-0.07$      & $-0.08^{*}$ & $-0.07^{*}$   & $-0.09^{*}$ & $-0.07^{*}$   & $-0.07^{*}$ \\
Edu: Basic        & $-0.05$      & $-0.04$      & $-0.04$     & $-0.06^{*}$   & $-0.04$     & $-0.06^{*}$   & $-0.06^{*}$ \\
Economic Status: iOS users & $0.02$       & $0.02$       & $0.02$      & $0.03$        & $0.02$      & $0.03$        & $0.03$ \\
Ln(Geographic Distance) & $-0.05$      &              &           &              &           &              &              \\
Ln(GDP Difference)      &              & $0.81^{*}$   &           &              &           &              &              \\
Ln(GDP/capita Difference) &            &              & $\textcolor{blue}{0.33^{***}}$ &              &           &              &              \\
Ln(Power Distance Difference) &           &              &           & $\textcolor{blue}{0.65^{***}}$  &           & $\textcolor{blue}{0.60^{***}}$  &              \\
Ln(Individualism Difference) &            &              &           & $0.97^{**}$   &           & $\textcolor{blue}{0.77^{***}}$  &              \\
Ln(Motivation towards Achievement and Success Difference) & & &           & $\textcolor{blue}{1.07^{***}}$  &           & $\textcolor{blue}{0.97^{***}}$  &              \\
Ln(Uncertainty Avoidance Difference) &     &              &           & $\textcolor{blue}{0.79^{***}}$  &           & $\textcolor{blue}{0.91^{***}}$  &              \\
Ln(Long Term Orientation Difference) &     &              &           & $-0.43$       & $0.34^{*}$ &              &              \\
Ln(Indulgence Difference)  &              &              &           & $\textcolor{blue}{-1.29^{***}}$ &           & $\textcolor{blue}{-1.18^{***}}$ &              \\

Country Dummies     &              &              &           &              &           &              & Included \\
\midrule
$R^2$             & $0.07$       & $0.13$       & $0.07$      & $0.57$        & $0.07$      & $0.57$        & $0.57$ \\
Adj. $R^2$        & $0.04$       & $0.10$       & $0.04$      & $0.55$        & $0.04$      & $0.55$        & $0.54$ \\
Max. VIF value    & $1.01$       & $1.07$       & $1.02$      & $22.55$       & $1.01$      & $4.91$        & $1.14$            \\
Num. of Obs.      & $214$        & $214$        & $214$       & $214$         & $214$       & $214$         & $214$ \\
\bottomrule
\end{tabular}
\caption{Reports the OLS regression analysis regarding the \textbf{Effects of Distances on Culture Retention of Immigrants}. Here, base categories for demographic variables for all regression models are Age: 18-34, Gender: Female, Edu: Not disclosed, and OS: Android Users. \textbf{Dependent Variables: \text{CR$_1$}}. Explanatory variables are differences between the USA and our 8 countries across different dimensions of distances, i.e., Geographic, Economic, and Hofstede’s 6 Cultural dimensions. Significance level: $^{***}p<0.001$; $^{**}p<0.01$; $^{*}p<0.05$. \textbf{Robustness tests}: \textcolor{blue}{Bonferroni Correction} reconfirms the OLS-based regression analysis at $p<0.05$ for the \textcolor{blue}{variables color-coded in blue}.}
\label{tab:CR1_Country_Distances}
\end{center}
\end{table*}

\begin{table*}[ht!]
\scriptsize
\begin{center}
\begin{tabular}{l c c c c c c c}
\toprule
 & \multicolumn{1}{c}{Model 1} & \multicolumn{1}{c}{Model 2} & \multicolumn{1}{c}{Model 3} & \multicolumn{1}{c}{Model 4} & \multicolumn{1}{c}{Model 5} & \multicolumn{1}{c}{Model 6} & \multicolumn{1}{c}{Model 7} \\
\midrule
Intercept         & $\textcolor{blue}{0.96^{***}}$ & $-5.40$     & $-1.37^{**}$ & $-3.35^{**}$  & $-0.33$     & $\textcolor{blue}{-4.06^{***}}$ & $\textcolor{blue}{0.57^{***}}$ \\
Age: 35 to 55     & $0.01$       & $0.02$      & $0.02$       & $0.03$        & $0.02$      & $0.03$        & $0.03$ \\
Age: 56 and above & $0.02$       & $0.02$      & $0.04$       & $0.03$        & $0.02$      & $0.03$        & $0.03$ \\
Male Dummy        & $-0.04^{*}$  & $-0.04^{*}$ & $\textcolor{blue}{-0.05^{**}}$ & $-0.05^{***}$ & $-0.04^{*}$ & $\textcolor{blue}{-0.05^{***}}$ & $\textcolor{blue}{-0.05^{***}}$ \\
Edu: Advanced     & $-0.04^{*}$  & $-0.04$     & $-0.03$      & $-0.03$       & $-0.04$     & $-0.03$       & $-0.03$ \\
Edu: Basic        & $-0.02$      & $-0.02$     & $-0.02$      & $-0.02$       & $-0.02$     & $-0.02$       & $-0.02$ \\
Economic Status: iOS users & $0.01$       & $0.01$      & $0.01$       & $0.01$        & $0.01$      & $0.01$        & $0.01$ \\
Ln(Geographic Distance) & $-0.03^{*}$  &           &            &             &           &             &             \\
Ln(GDP Difference)      &            & $0.36^{*}$  &            &             &           &             &             \\
Ln(GDP/capita Difference) &         &           & $\textcolor{blue}{0.18^{***}}$ &             &           &             &             \\
Power Distance Difference &          &           &            & $\textcolor{blue}{0.35^{***}}$  &           & $\textcolor{blue}{0.32^{***}}$  &             \\
Individualism Difference  &          &           &            & $0.53^{***}$  &           & $\textcolor{blue}{0.44^{***}}$  &             \\
Motivation towards Achievement and Success Difference & & & & $\textcolor{blue}{0.50^{***}}$  &           & $\textcolor{blue}{0.45^{***}}$  &             \\
Uncertainty Avoidance Difference  &  &           &            & $\textcolor{blue}{0.40^{***}}$  &           & $\textcolor{blue}{0.46^{***}}$  &             \\
Long Term Orientation Difference  &  &           &            & $-0.21$       & $0.21^{*}$  &             &             \\
Indulgence Difference              &  &           &            & $\textcolor{blue}{-0.66^{***}}$ &           & $\textcolor{blue}{-0.61^{***}}$ &             \\

Country Dummies     &              &              &           &              &           &              & Included \\
\midrule
$R^2$               & $0.07$       & $0.07$      & $0.13$       & $0.56$        & $0.08$      & $0.55$        & $0.56$ \\
Adj. $R^2$          & $0.04$       & $0.04$      & $0.10$       & $0.53$        & $0.05$      & $0.53$        & $0.53$ \\
Max. VIF value      & $1.01$       & $1.02$      & $1.07$       & $22.56$       & $1.01$      & $4.91$        & $1.14$        \\
Num. obs.           & $214$        & $214$       & $214$        & $214$         & $214$       & $214$         & $214$ \\
\bottomrule
\multicolumn{8}{l}{\scriptsize{$^{***}p<0.001$; $^{**}p<0.01$; $^{*}p<0.05$}}
\end{tabular}
\caption{\textbf{Robustness Analysis: Effects of Distances on Culture Retention of Immigrants}. Here, base categories for demographic variables for all regression models are Age: 18-34, Gender: Female, Edu: Not disclosed, and OS: Android Users. Dependent Variables: $\text{CR}_2$. Significance level: $^{***}p<0.001$; $^{**}p<0.01$; $^{*}p<0.05$. \textcolor{blue}{Bonferroni Correction} reconfirms the OLS-based regression analysis at $p<0.05$ for the \textcolor{blue}{variables color-coded in blue}.}
\label{tab:CR2_Country_Distances}
\end{center}
\end{table*}

\textbf{Demographic Characteristics}: Following prior studies, we have controlled for different demographic categories \cite{dubois2018studying, gil2019demographic, stewart2019rock}, i.e., we extract audience size for a specific demographic profile. Based on the literature, the following demographic attributes are considered for disaggregation: 

\textit{Age:} 18 to 34; 35 to 55; 56 and above [\textit{3 categories}] 

\textit{Gender:} Male and Female [\textit{2 categories}]

\textit{Education:} Basic (such as High School, Some High School, Undergrad, Associate Degree, etc.); Advanced (such as Professional Degree, Master’s Degree, Doctorate Degree, etc.); Undisclosed/unspecified [\textit{3 categories}] 

\textit{Economic Status:} iOS users vs.\ Android users [\textit{2 categories}] - prior studies argue that using an iOS, i.e., Apple, device to access Facebook is generally a proxy for higher wealth \cite{fatehkia2020mapping,fatehkiaetal22frontiers}.

\section{Analysis \& Findings}

\textbf{Data Snapshot:} Table \ref{Tab:data_snapshot} presents summary statistics of explanatory variables, including mean, standard deviation (S.D.), minimum/maximum values, and the corresponding countries with the lowest/highest scores. As anticipated, developed nations, such as Germany, demonstrate robust economic status and socio-political freedom. Additionally, the economic disparities compared to the USA are relatively minimal for Germany. Conversely, significant cultural gaps are observed, particularly between the USA and developing nations such as Vietnam, Mexico, and Bangladesh, where socio-political freedoms are also low.

\textbf{Statistical Analysis:} Tables \ref{tab:CR1_CountryCharacteristics} to \ref{tab:CR2_Country_Distances} report the findings of the Ordinary Least Squares (OLS) regression analysis. For each country, we have extracted audience estimates for 36 combinations: 3 (age) * 2 (gender) * 3 (education) * 2 (economic status). Ideally, we should have 288 observations (36 * 8 countries) for our analysis, but we have only 214 observations due to data sparsity - as evident from Figures 1 and 4. Dependent Variables for Tables 3 and 5 are \text{CR$_1$}.

\textit{\textbf{RQ1: Do Origin-Country Characteristics Matter?} }

Table \ref{tab:CR1_CountryCharacteristics} reports the impact of various country-specific characteristics on cultural retention among immigrants. Model 1 is our base model with only demographic or control variables. We note that the male gender exhibits a consistent negative association with culture retention across all models. Similarly, advanced education shows a primarily negative association with culture retention across models, although the significance varies for both variables. Intuitively, male and highly educated immigrants have a higher propensity for social interactions, and this, in turn, negatively impacts their cultural retention tendencies, and \textit{refutes the mosaic hypothesis}. Notably, Model 1 has very low explanatory power (i.e., Adj. R$^2$ is 0.03), compared to other models in this table. It strongly indicates that demographic factors (e.g., age, gender, education, or economic wealth) have a limited impact on cultural retention. 

Next, we find Political rights, Civil liberties, and Linguistic Diversity emerge as crucial determinants, exhibiting statistically significant ($p<0.001$) negative associations with culture retention. Intuitively, these findings highlight the importance of political or social freedoms as well as linguistic diversity in the native country in diminishing cultural retention among immigrants, and \textit{our analysis refutes the mosaic hypothesis.}
Conversely, factors such as the presence of other immigrants from the native country in the host country demonstrate statistically significant ($p<0.001$) positive relationships, indicating their impact in fostering cultural retention within immigrant communities, and \textit{supports the mosaic hypothesis} - a vibrant presence of diaspora allows an immigrant to find his own community in the host country and retain his native culture. Surprisingly, we didn't find any statistically significant relationship between the economic health  (i.e., GDP and GDP per capita) or HDI of a country and the cultural retention of immigrants. Similarly, we anticipated that exposure to international tourists in the native country would negatively impact cultural retention, but we didn't get a statistically significant relationship. Probably, international tourists go to a few specific tourist destinations or popular cities in a country. So, our variable fails to capture the exposure to diverse cultures. For ease of interpretation, we visually present all model findings together in {Figure \ref{fig:effects_country_char_dist}}. 

Lastly, the Adj. R$^2$ values are modest across models, but this is typical for sociocultural studies using social media data. Importantly, our aim is not to develop a predictive model, where Adj. R$^2$ would be a primary evaluation metric; rather, our focus is on the direction and significance of theoretically driven relationships.

\begin{figure}[ht!]
    \centering
    \includegraphics[scale=0.15]{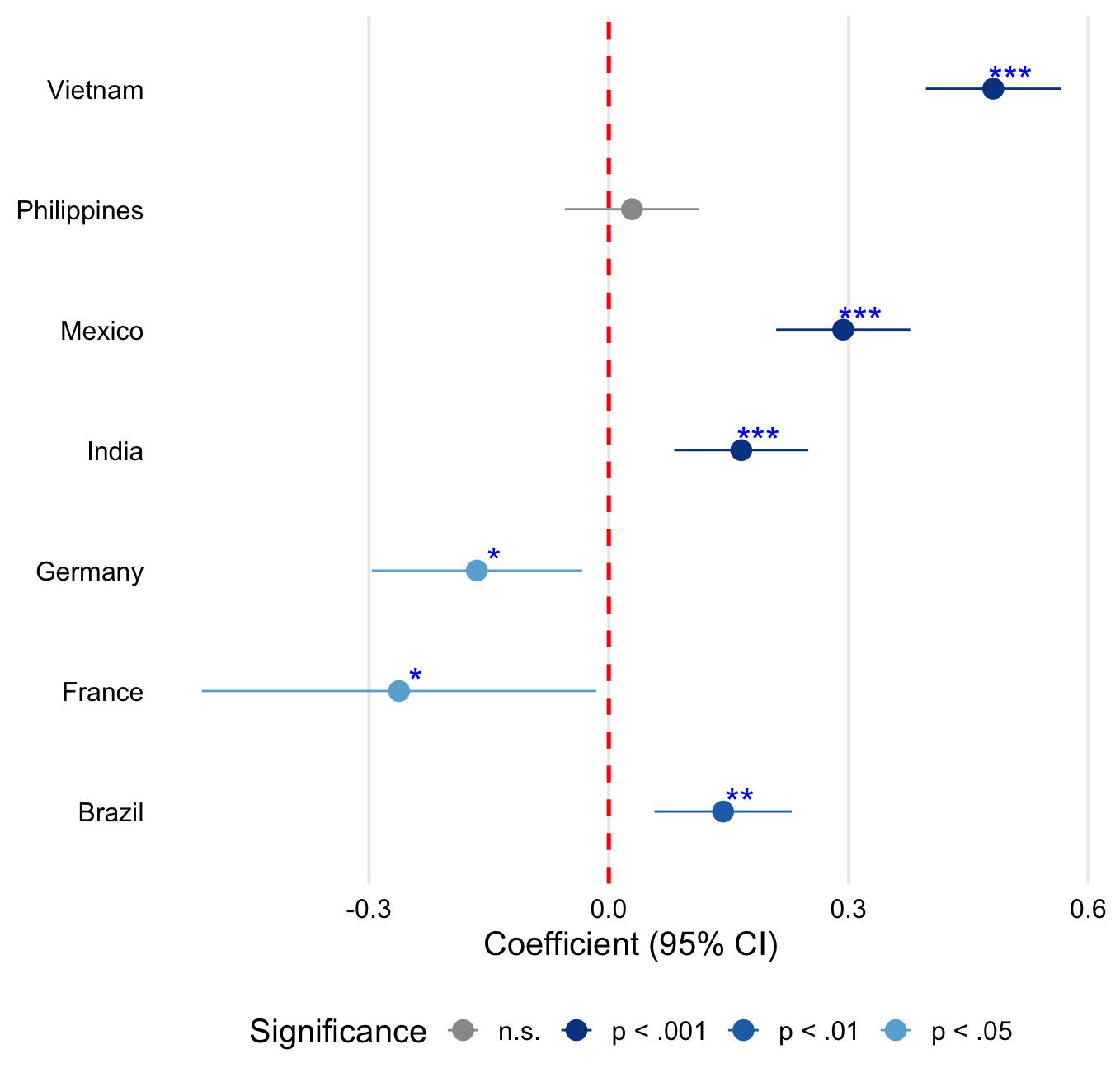}
    \caption{Visual representation of country-level fixed effects on cultural retention (CR$_1$), with Bangladesh as the base category. The \textcolor{cyan}{cyan (low statistical significance)} and \textcolor{blue}{blue (high statistical significance)} circles represent the coefficients of country dummy variables estimated in Model M7 (Table \ref{tab:CR1_Country_Distances}). Horizontal lines indicate 95\% confidence intervals, and \textcolor{blue}{blue stars} denote significance levels.
}
    \label{fig:country_level_coeff}
\end{figure}

\begin{figure}[ht!]
    \centering
    \includegraphics[scale=0.17]{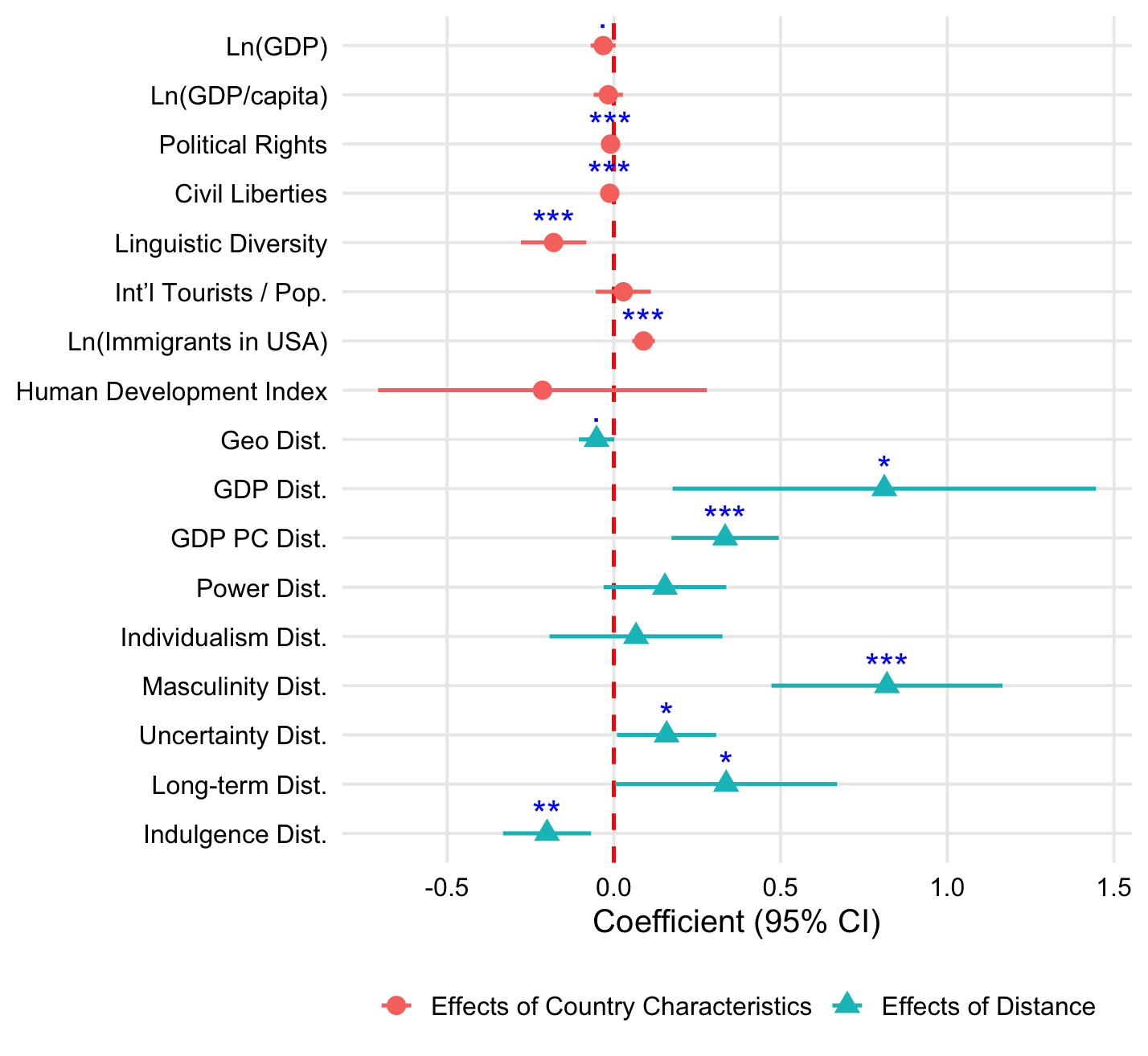}
    \caption{Visual representation of the fixed effects of country characteristics (e.g., GDP, HDI) and distance dimensions (e.g., economic, cultural) on cultural retention (CR$_1$). The estimated coefficients from regression models in Table \ref{tab:CR1_CountryCharacteristics} are shown as \textcolor{red}{red circles }(country characteristics), and those from Table \ref{tab:CR1_Country_Distances} are shown as \textcolor{cyan}{cyan triangles }(distance dimensions). Horizontal lines indicate 95\% confidence intervals, and stars denote conventional significance levels ($^{*}p < 0.05$; $^{**}p < 0.01$; $^{***}p < 0.001$). Estimates are standardized for comparability across models.}
    \label{fig:effects_country_char_dist}
\end{figure}

\textit{\textbf{RQ2: Does Host–Native Distance Matter?}}

In Table \ref{tab:CR1_Country_Distances}, Model 1 incorporates the geographic distance (between the USA and native countries of immigrants), and we didn't find any significant impact. Next, we consider economic distance, i.e., differences between the USA and respective native countries in terms of GDP and GDP per capita, and find a statistically significant positive effect. It is worth noting that the direct impact of these variables was insignificant in Table \ref{tab:CR1_CountryCharacteristics}. Thus, it can be argued that the propensity of cultural retention is high when the gap between the GDP/capita (or affordability) of immigrants and host country citizens is large. Next, we consider cultural distance and incorporate Hofstede’s 6 cultural dimensions, but we find multicollinearity issues (i.e., disproportionately high VIF value). Hence, to avoid multicollinearity - based on VIF analysis, Model 5 considers only Long Term Orientation Difference and reports a statistically significant ($p<0.05$) positive effect (with the highest VIF value of 1.01), and Model 6 reports the other 5 cultural dimensions (with the highest VIF value of 4.91). As anticipated, we find that higher cultural distance leads to higher cultural retention for 4 dimensions in Model 6 (with $p<0.001$) - except for Individualism. A negative impact of Individualism indicates immigrants tend to move away from a restrained culture to a more relaxed attitude toward life and embrace the new culture. Notably, Adj. R$^2$ significantly improved from 0.03 (in Model 1 of Table \ref{tab:CR1_CountryCharacteristics}) to 0.55 (in Model 6 of Table \ref{tab:CR1_Country_Distances}). This improvement in Adj. R$^2$ strongly confirms the impact of cultural distance on cultural retention. In short, \textit{economic and cultural distances broadly support the mosaic hypothesis.}

It is possible that we may have ignored other determinants of cultural retention, and our list of variables is certainly not exhaustive. In response to potential omitted variable bias, we have controlled for country effects using dummy variables and found that most country dummies are significant, with a relatively high Adj. R$^2$ of 0.56. For brevity and ease of interpretation, we visually report these country effects in  {Figure \ref{fig:country_level_coeff} }.

\textbf{Visual Representations of OLS Analysis:} {Figures \ref{fig:country_level_coeff} and \ref{fig:effects_country_char_dist}} visually depict the findings of our OLS analysis. {Figures \ref{fig:country_level_coeff} answers - does country matter? From Model 7 of Table \ref{tab:CR1_Country_Distances}}, the country coefficients and their significance levels are reported in {Figure \ref{fig:country_level_coeff}}, with Bangladesh as the base category. We find that cultural retention scores are significantly lower for Germany and France, two developed countries in our sample, compared to other developing countries, indicating that \textit{the country of origin does matter}. Next, we plot the effects of country characteristics (through \textcolor{red}{red circles}) and various dimensions of distance (through \textcolor{cyan}{cyan traingles}) in {Figure \ref{fig:effects_country_char_dist}}, based on the findings from {Tables \ref{tab:CR1_CountryCharacteristics} and \ref{tab:CR1_Country_Distances}}. {Figure \ref{fig:effects_country_char_dist}} strongly suggests that a greater distance is associated with higher cultural retention. For instance, economic distance, captured through GDP difference, has the highest impact on cultural retention. While country context matters and most results are statistically significant, their impacts are relatively small compared to distance.

\textbf{Robustness Analysis:} For the sake of robustness, we consider an alternate operationalization of Cultural Retention, CR$_2$, as the natural logarithm of our original CR$_1$. 

\begin{equation*}
\text{CR$_2$} = Ln\frac{\text{$CI(Imm)_{\text{Ctry}}$}}{\text{$CI(Nat)_{\text{Ctry}}$}}
\end{equation*}

In {Tables \ref{tab:CR2_CountryCharacteristics} and \ref{tab:CR2_Country_Distances}}, we repeat our OLS regression analysis with \text{CR$_2$} - as the dependent variable. It is reassuring to note that our results remain consistent for all variables except Geographic Distance. Previously, we didn't find a statistically significant result for \textit{Ln(Geographic Distance)}, but in our robustness analysis, we find a negative and weakly significant effect on cultural retention, \textit{refuting the mosaic hypothesis} - as anticipated. 

The \textcolor{blue}{\textbf{Bonferroni correction}} is a statistical method used to reduce the risk of falsely identifying a result as significant when performing multiple tests on the same dataset. Our Bonferroni correction reaffirms the results for most key variables - these variables are \textcolor{blue}{color-coded in blue} in {Tables \ref{tab:CR1_CountryCharacteristics}, \ref{tab:CR1_Country_Distances}, \ref{tab:CR2_CountryCharacteristics} and \ref{tab:CR2_Country_Distances}}.

\section{ Limitations and Future Scope}
This study has a few limitations that offer potential avenues for future research. \textit{First}, our analysis is based on cross-sectional data. Facebook data didn’t allow us to consider how long individuals have been in the host country. For instance, cultural retention might be stronger during the initial years of migration but could diminish as immigrants integrate into the host society. Without longitudinal data, it is challenging to identify such temporal dynamics. \textit{Second}, cross-lingual and cross-country comparisons of Facebook interest data have some inherent limitations. For instance, the number of interests available on the platform might vary by country, making normalization of interests across countries/cultures difficult - as reported in {Figure \ref{fig:grouped Interests}}. \textit{Third}, audience sizes for specific interests may fluctuate over time, raising concerns about the stability of findings. For instance, the timing of a popular cricket (sports) tournament in India, such as the Indian Premier League (IPL), may impact the Indian audience size for cricket-related interests. \textit{Fourth}, the lack of fine-grained sub-national data, as reported in {Figures \ref{fig:Geographic distribution} and \ref{fig:Own-culture engagement}}, limits the study's ability to capture local cultural integration vis-à-vis retention dynamics. Cultural assimilation or retention processes can be influenced by factors such as immigrant density or political leanings (e.g., Republican vs. Democratic regions). We tried to capture immigration density at the country level, but sub-national control could have enhanced the rigor of our analysis. 
\textit{Lastly}, we are not sure whether low cultural retention supports the melting pot hypothesis. According to \citet{berry1997immigration}, low cultural retention might not always lead to separation, rejecting the host culture to retain one’s own, but could instead result in marginalization, disengaging from both cultures. Conversely, high cultural retention might not imply assimilation, embracing the host culture over the native one, but could reflect integration, embracing both cultures. The scope of this paper did not allow us to examine these finer nuances, but future studies can explicitly investigate these aspects.

In short, while our study provides valuable insights, such as elucidating the potential of digital data to understand intricate issues like online cultural retention of immigrants, at the same time, the nature of the data restricts us from performing a fine-grained analysis. This implicitly indicates the relevance of a mixed-method collaborative approach, where information science researchers should collaborate with qualitative or survey-based researchers to get a holistic understanding of complex societal issues.

\section{Conclusion}
As immigrants navigate the unfamiliar terrain of host countries, they often struggle with their native identity. 
In this context, the issue of retaining one's native identity (or cultural interests of one's native country) is multifaceted and influenced by a variety of factors. We have explored country-specific characteristics and the distances, geographic, economic, and cultural, between native and host countries to investigate their impact on online cultural retention among immigrants. Our study highlights the significance of both sets of factors, with country-specific variables such as political rights, civil liberties, and linguistic diversity displaying a negative effect on cultural retention. Conversely, the distances between countries, particularly economic and cultural differences, positively impact the cultural retention of immigrants, \textit{aligning with the mosaic hypothesis}. Our study discards a one-size-fits-all approach because integration outcomes depend not only on immigrants’ choices but also on structural factors of the host and native countries - a key insight for policymakers striving to support inclusivity within host countries.

\textbf{Ethical Considerations: } Facebook’s API allows the extraction of anonymous and aggregate-level user data, and it has no re-identification risk for individual users. As we are struggling with issues of integration and cultural identity, this study on cultural retention can potentially enrich our appreciation of diverse cultural practices.

\section{Acknowledgments}
This work was partially supported by the Wallenberg AI, Autonomous Systems and Software Program (WASP) funded by the Knut and Alice Wallenberg Foundation.

\bibliography{aaai25}

\end{document}